\documentclass[12pt,preprint]{aastex}
\usepackage{graphicx}
\usepackage{psfig}
\usepackage{mathrsfs}

\shorttitle{VFTS 682 \& Dark GRB Progenitor} \shortauthors{Zhang \& Stanek}

\begin{document}

\title{The Very Massive and Hot LMC Star VFTS 682: Progenitor of a Future Dark Gamma-Ray Burst?}
\author{Dong Zhang\altaffilmark{1} and K. Z.~Stanek\altaffilmark{1,2}}

\altaffiltext{1}{Department of Astronomy, The Ohio State University,
140 W. 18th Ave., Columbus, OH, 43210; dzhang, kstanek@astronomy.ohio-state.edu}

\altaffiltext{2}{Center for Cosmology \& Astro-Particle Physics, The
Ohio State University, 191 West Woodruff Ave., Columbus, OH, 43210}

\begin{abstract}

VFTS 682, a very massive and very hot Wolf-Rayet (WR) star recently
discovered in the Large Magellanic Cloud near the famous star cluster
R136, might be providing us with a glimpse of a missing link in our
understanding of Long Gamma-Ray Bursts (LGRBs), including dark
GRBs. It is likely its properties result from chemically homogeneous
evolution (CHE), believed to be a key process for a massive
star to become a GRB. It is also heavily obscured by dust extinction,
which could make it a dark GRB upon explosion. Using {\em Spitzer}\/
data we investigate the properties of interstellar dust in
the vicinity of R136, and argue that its high obscuration is not
unusual for its environment and that it could indeed be a slow runaway
(``walkaway'') from R136. Unfortunately, based on its current mass loss
rate, VFTS 682 is unlikely to become a GRB, because it will lose too
much angular momentum at its death. If it were to become a GRB, it probably would also
not be dark, either escaping or destroying its surrounding dusty region.
Nevertheless, it is a very interesting star, deserving further studies, and
being one of only three presently identified WR stars (two others in the Small Magellanic Cloud)
that seems to be undergoing CHE.

\end{abstract}

\keywords{stars: mass-loss --- stars: Wolf-Rayet --- gamma-ray burst: general --- Magellanic Clouds}

\section{Introduction}

Ideally, in order to fully understand the mapping between massive star
progenitors and when and how they explode (or not,
\citealt{Kochanek08}), we would have extensive multi-wavelength data
obtained for many such explosions both BEFORE and AFTER the event. The
AFTER part has certainly undergone an explosive growth in the last
decade or so, due to many successful supernova (SN) searches such as the galaxy-targeted {\em Lick Observatory Supernova Search} (LOSS), the {\em Catalina Real-time Transient Survey} (CRTS), the {\em Robotic Optical Transient Search Experiment} (ROTSE), the {\em Palomar Transient Factory} (PTF); and gamma-ray burst searches such as the {\em High Energy Transient Explorer-2 (HETE-2)} and {\em Swift}.  The BEFORE part is naturally limited by the fact that massive stars, while relatively bright, are only observable in the local Universe ($d\lesssim10\;$Mpc). Here we have had to rely on proximity (SN 1987A, \citealt{SN1987A}),
luck based on archival data (e.g., \citealt{Smartt09}), or the first systematic
campaign to monitor future SN progenitors (\citealt{Szczygiel11}),
where many nearby galaxies are observed to sufficient depth so eventually they will provide progenitor data for a
significant number of future SNe. We have by now compiled a fair amount of
information on SN progenitors, such as the blue supergiant progenitor for SN 1987A
(\citealt{White87}) or dusty progenitors of SN 2008S and 2008 NGC 300
transient (\citealt{Prieto08}; \citealt{Prieto08b}).

However, for very rare explosive events, such as long-duration
gamma-ray bursts (LGRBs), the systematic approach that works for the
normal core-collapse SNe is not yet possible because the events are so rare. 
Therefore the prospect of future systematic observational efforts that could identify nearby LGRB progenitors is very dim. Here we have to rely on a more extended chain of
reasoning, with data being supplemented with reasonable, theoretical guesses. Discovery of the connection between LGRBs and
broad-lined Type Ic SNe (e.g., \citealt{Stanek03};
\citealt{Hjorth03}), thought to result from core-collapse of hydrogen-free
massive stars, favors two possible progenitor models: single massive
Wolf-Rayet (WR) stars with rapidly rotating cores (\citealt{FS85}), or lower mass
helium stars stripped by a close binary companion (\citealt{Podsiadlowski04}). The prompt emission of an initial LGRB, lasting seconds or minutes, is
usually followed by a multi-wavelength afterglow which lasts days to even years.
Although X-ray afterglows of LGRBs are nearly always detected by {\em Swift} and {\em Fermi}, detection of optical and infrared
afterglows is less common. A dark GRB is defined by either an absent or faint optical afterglow relative to its X-ray emission.
The rapid detection of X-ray afterglows with {\em Swift} revealed that the dark
fraction of LGRBs is about $30\%$ of GRBs (\citealt{AS07};
\citealt{Perley09}). In general, besides the intrinsically optically faint
GRBs, the optical attenuation of dark GRBs can be caused by dust extinction in GRB
host galaxies, foreground extinction, or Lyman-$\alpha$ absorption
by neutral hydrogen at high redshifts (\citealt{Perley09}).

Study of massive stars in very nearby galaxies can supply the
other missing links in our understanding of LGRBs.  The recently
uncovered very massive stars up to $300M_{\odot}$ in the Large
Magellanic Cloud (LMC) young cluster R136 extend our
knowledge of massive star formation and evolution
(\citealt{Crowther10}).  In particular, the newly discovered very
massive WR star VFTS 682, located 30 pc away from R136, drew our attention,
mainly because of its very high foreground dust extinction
and possibility of being a GRB progenitor (\citealt{VFTS682}). Its
unusually high effective temperature can be understood as the consequence of chemically homogeneous evolution
(CHE), which was proposed as the crucial part of the process of
producing LGRBs (\citealt{Yoon05}). If VFTS 682 will
indeed make a GRB at the end of its evolution, the high
foreground dust extinction could mean that it will
become a dark GRB. However, its high mass and strong mass-loss seems
to prevent it from being a GRB progenitor, and the potential afterglow might destroy the surrounding dust even if VFTS 682 eventually produces a GRB. Therefore, it is worthwhile to look into the details of VFTS 682's evolution and the fate of its circumstellar environment. Section \ref{Data}
describes the dust properties in the VFTS 682 vicinity region. Section
\ref{Progenitor} investigates the possibility of VFTS 682 producing a
LGRB and Section \ref{Dark} discusses the fate of dusty clouds around VFTS 682. Conclusions are presented in
Section \ref{Con}.

\section{Investigating the Properties of Interstellar Dust in the Vicinity of VFTS 682}\label{Data}

\begin{figure*}[p]
\epsscale{1.0}
\plotone{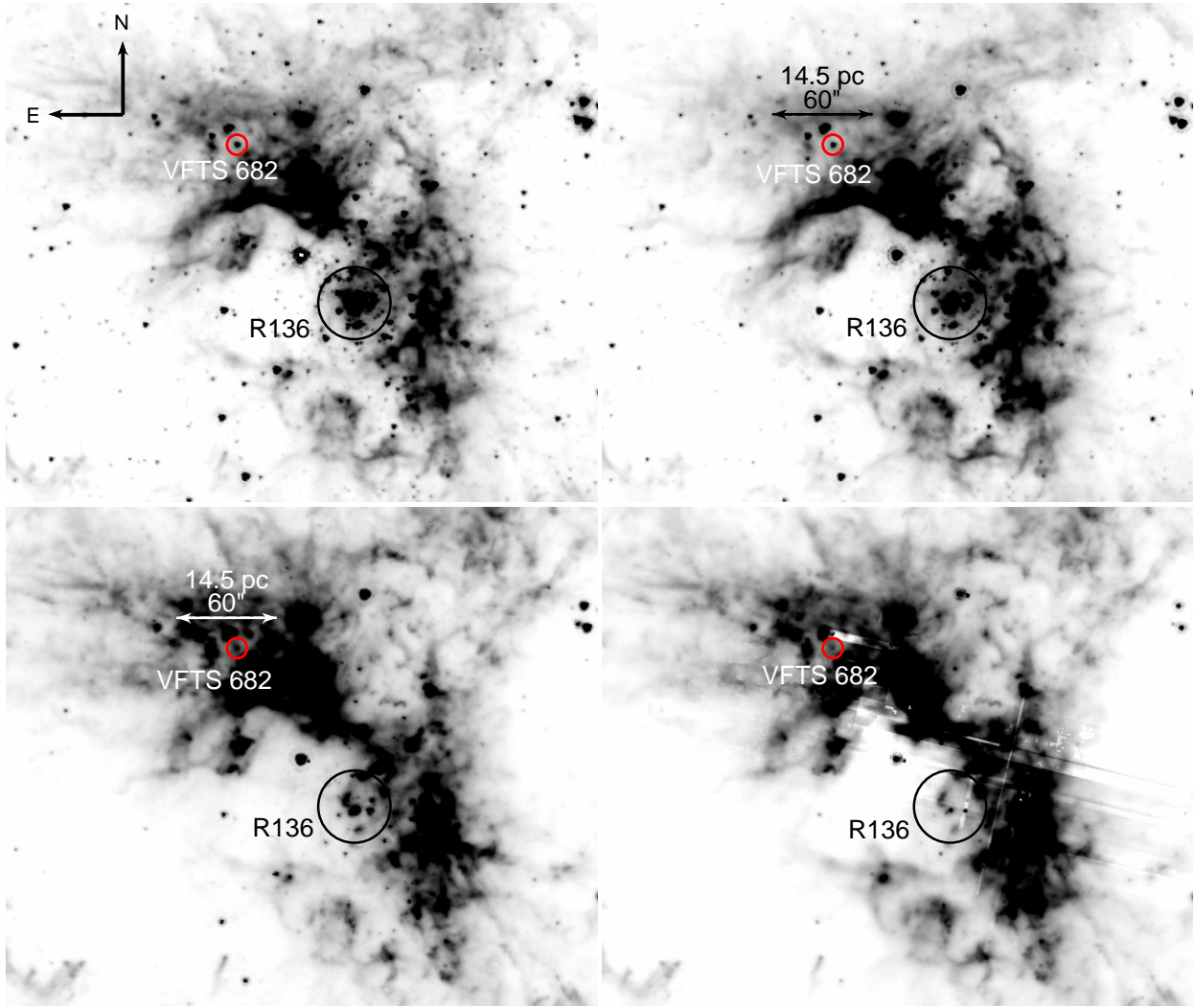}
\caption{{\em Spitzer}-IRAC archival images of the 30 Doradus region
in 3.6, 4.5, 5.8 and 8.0 $\mu$m (upper left to bottom right) IRAC
photometric bands. The location of VFTS 682 is marked with the small
circle, while the location of the R136 cluster is shown with the
large circle. The 60$\arcsec$ scale bar corresponds to roughly
$14.5\,$pc for an assumed LMC distance of $50\,$kpc.}
\label{fig1}
\end{figure*}

\begin{figure*}[p]
\epsscale{1.0}
\plotone{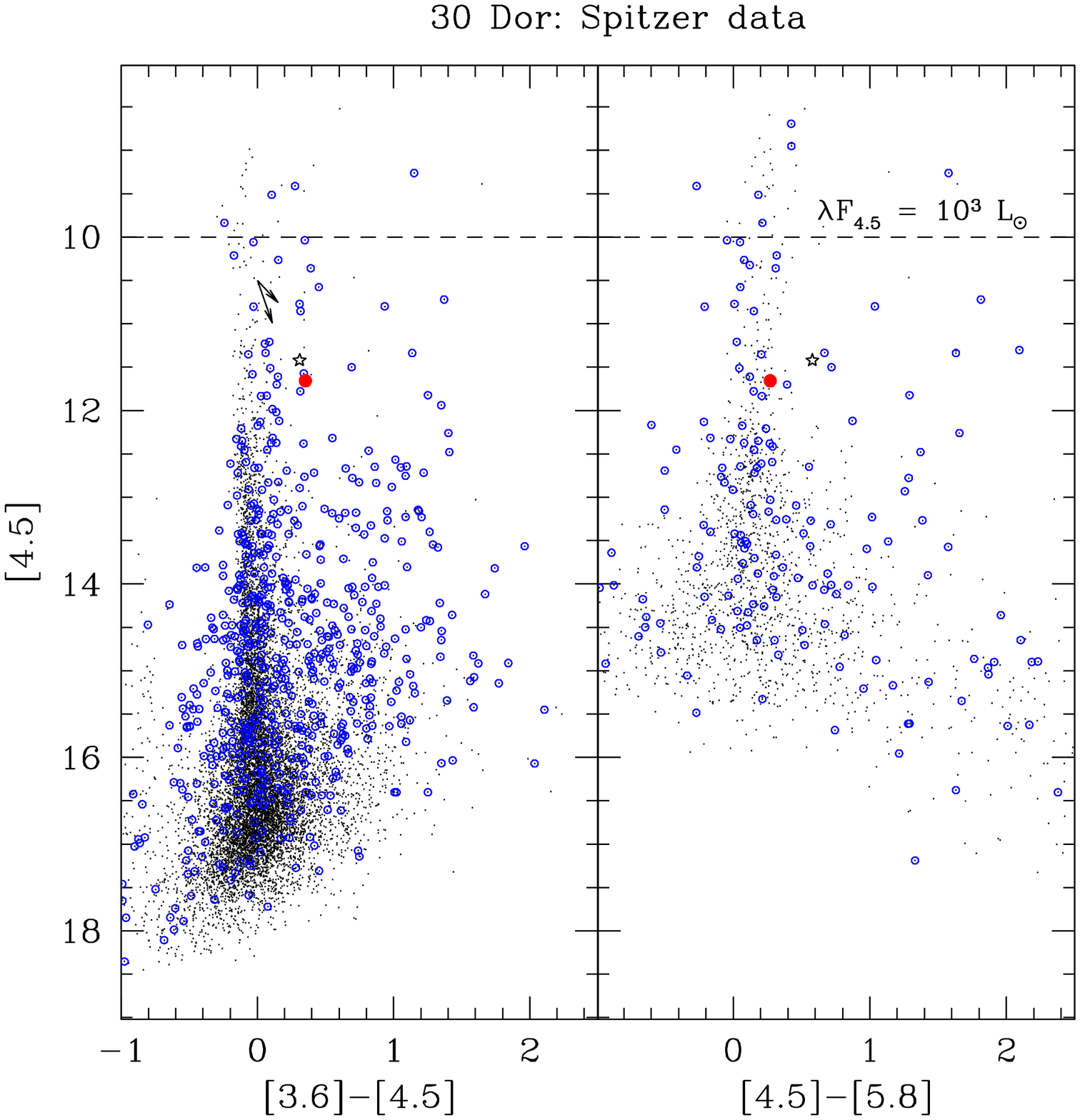}
\caption{Mid-IR color-magnitude diagrams for the vicinity of VFTS 682
obtained using Daophot on the archival {\em Spitzer} IRAC images of $[4.5], [3.6]-[4.5]$ CMD (left), and $[4.5], [4.5]-[5.8]$ CMD (right).  The small dots show all matched point sources in the $20\arcmin \times 20\arcmin$ field, while
the bigger open circles represent sources within $3\arcmin$ (roughly
$44\;$pc) of R136. The red filled dot represents our photometry for
VFTS 682, while the open star shows the values as measured by the SAGE
project (Meixner et al.~2006). Also shown in the left panel are two reddening vectors corresponding to $A_K$ value of
1.0 (about twice as high as derived by Bestenlehner et al.~2011 for
VFTS 682), for two different values of the interstellar reddening law,
$R_V=3.1$ and 4.7. For discussion see Section~\ref{Data}.}
\label{fig2}
\end{figure*}

Among the many striking properties of VFTS 682 discussed in
\cite{VFTS682}, the high value they derived for the foreground
extinction to that star, $A_V=4.45 \pm 0.12$, particularly drew our
attention. Indeed, for a random line of sight to an object in the LMC,
a typical value of the interstellar extinction would be much lower,
$A_V\sim 0.3$ (e.g., \citealt{{Pejcha09}}).
At the same time, other very massive, nearby
stars in R136---the central cluster of the 30 Doradus region---have
estimated interstellar extinctions near $A_V\sim 2.0$ (see Table 3
in \citealt{Crowther10}), significantly lower than VFTS 682. This high
value of extinction could provide an interesting clue to the
origin of VFTS 682, because if such values were indeed very rare in
the 30 Dor region, it would argue against VFTS 682 being a slow
runaway (``walkaway'') from the R136 cluster.

To investigate the properties of interstellar dust extinction in the
vicinity of VFTS 682, we analyzed the archival {\em Spitzer
Space Telescope} IRAC data for the 30 Dor region, which were obtained in 2003 by the ``Comparative
Study of Galactic and Extragalactic HII Regions'' program (PI: Houck;
Program ID: 63). For the purpose of this investigation, we limited our
analyses to a region $20\arcmin \times 20\arcmin$ centered on R136.

In Figure~\ref{fig1} we show $3.6, 4.5, 5.8$ and $8.0\;\mu$m IRAC
images of the region near VFTS 682 and 30 Doradus.
We see a progression of still seeing significant
stellar light in the $3.6\mu$m band, so the cluster still features
prominently in this band, to being completely dominated by warm dust
emission and PAH emission in the $8.0\mu$m band, where the cluster
virtually disappears.  If we can use the $8.0\mu$m band emission as a
proxy for dust column density, then the high value of $A_V$ towards
VFTS 682 is not at all unusual in the 30 Dor region, and there should
be lines of sight with 5 to 10 times higher values of interstellar
extinction near the location of VFTS 682.  Therefore, despite our
initial expectation, the high value of $A_V$ does not bring any new
information about the nature of VFTS 682, namely was it ejected from
R136 or was it born in situ.  However, the presence of such large amounts
of spatially complex dust raises the interesting possibility of
there being other stars like VFTS 682 in the vicinity of R136,
that are hidden behind still more dust.

For a more quantitative analyses of the IRAC data, we used Daophot
(\citealt{Stetson92}) to identify point sources and measure their
fluxes in the $3.6, 4.5$ and $5.8\;\mu$m images. In Figure~\ref{fig2}
we show the resulting mid-IR color-magnitude diagrams (CMDs).
In the $[4.5], [3.6]-[4.5]$ CMD we more or less agree with SAGE that VFTS 682 has a small $4.5\mu$m flux excess,
but we find a bluer $[4.5]-[5.8]$ color, most likely due to differing treatments of
the extended dust emission near VFTS 682. In any case, the amount of
resolved mid-IR emission at the position of VFTS 682 is very small
compared to its total bolometric luminosity.  It is clear from these
CMDs that the mid-IR properties of VFTS 682 are not at all unusual
compared to other stars in the 30 Dor region, and there are many other
stars nearby that are either more obscured or have significantly more
mid-IR emission around them.

\section{Can VFTS 682 be a GRB Progenitor?}\label{Progenitor}

VFTS 682 has a high temperature $T_{\rm eff}=52.2\pm2.5$ kK and
luminosity $\log(L/L_{\odot})=6.5\pm0.2$, placing it blueward of the
zero-age main sequence in the Hertzsprung-Russell (HR) diagram, which
can be explained by CHE. Since CHE is believed to be the key process to
make a LGRB, \cite{VFTS682} suggest VFTS 682 as a possible LGRB progenitor. A key ingredient to maintaining a CHE is
sufficiently fast stellar rotation to induce a complete chemical
mixing (\citealt{Schwarz58}). However, the problem is that CHE might
cease when the strong mass-loss from VFTS 682 carries away too much
angular momentum before its death.

Theoretically low metallicity is favorable for single stars to be GRB
progenitors, mainly because low metallicity leads to low mass-loss rates,
that sustain the fast rotation required by LGRB models (\citealt{Yoon05}; \citealt{WoosHeg06}). Recent simulations
proposed a metallicity threshold of $Z\leq0.004$ for
GRB production (\citealt{Yoon06}). Observations find that local LGRBs
associated with supernovae (SNe) have metal-poor host galaxies with
oxygen abundance of $12+\log\rm{(O/H)}<8.6$, or $Z<(0.2-0.5)Z_{\odot}$ (\citealt{Stanek06};
\citealt{Modjaz08}), but not as metal-poor as required by models. \cite{Stanek06} argued that LGRBs trace only
low-metallicity star formation. Although one or
two high-metallicity GRB hosts have been discovered
(\citealt{Levesque10}), all of these GRBs lack an SN signature,
leaving it unclear whether those GRBs can be
linked with star formation or not. The half-solar metallicity of VFTS
682 is just in the upper metal range given by LGRB-SN events,
so it is unclear whether it can be a LGRB
progenitor just based on the metallicity criterion. A more direct constraint on the fate of
VFTS 682 can be derived from its other physical parameters as follows.

The observed present mass-loss rate of VFTS 682 is $\log(\dot{M}/M_{\odot}$
yr$^{-1})=-4.4\pm0.2$, with an estimate present-day mass as
$M\sim150M_{\odot}$ (\citealt{VFTS682}). The mass-loss timescale for
VFTS 682 of $\tau_{\rm wind}=M/\dot{M}\sim3.8$ Myr is comparable to
the stellar nuclear timescale $\tau_{\rm nuc}\sim5\times10^{9}$ yr
$(M/M_{\odot})(L/L_{\odot})^{-1}\sim2.6$ Myr for the current hydrogen
abundance $X_{H}=0.55$. This means that VFTS 682 can easily be stripped
given its current mass-loss rate. For simplicity, if we take a remaining
lifetime for VFTS 682 on the main-sequence (MS) as $\sim2$ Myr with a
constant mass-loss rate as its present value, the total mass lost
when it evolves off the MS will be $M_{\rm loss}\sim80M_{\odot}$. Since the hydrogen abundance
will continue to decrease from its current value, while
the luminosity $L$ changes a little with both deceasing
$M$ and $X_{H}$ (\citealt{Grafener11}), the theoretically predicted
mass-loss rate $\dot{M}\propto10^{-0.45X_{H}}(L/L_{\odot})^{0.42}$
increases in the future (\citealt{Grafener08}). If the hydrogen abundance smoothly
drops from $X_{H}=0.55$ to $X_{H}\simeq 0$, the total mass-loss increases from the rough estimate $\sim80M_{\odot}$
to as high as $\sim100M_{\odot}$. This would leave a remaining core $M_{r}\sim70-50M_{\odot}$ star as the star leaves the MS,
and its subsequent evolution could eject even more mass. Unlike normal stars, which include core and envelope
components, stars undergoing CHE can be approximated as a chemically mixed, rigidly rotating bodies (\citealt{MM00}).
Treating VFTS 682 as a rigid object with a current surface rotation velocity
$v_{0}$ and radius-mass relation $R\propto
M^{\alpha}$, the final rotation velocity $v_{f}$ at the end of MS is
$v_{f}/v_{0}=(M_{r}/M_{0})^{(3-2\alpha)/2}$. Taking $\alpha\sim1$ for
very massive stars (\citealt{Yungelson08}), we have
$v_{f}/v_{0}\sim0.6-0.7$ and the star will have lost more than 85\% of its
current angular momentum before its death. For currently $v_{0}\leq700$
km s$^{-1}$, the stellar final rotation velocity $v_{f}\leq$500 km
s$^{-1}$ is unlikely high enough to maintain CHE in the LMC (see the criterion in \citealt{MM00}, or Fig. 7 in
\citealt{Brott11}). VFTS 682 will have a hydrogenic envelope at death, thus cannot become a LGRB.
On the other hand, if VFTS 682 has an incredibly rapid rotation velocity $v_{0}>700$ km s$^{-1}$, the
possibility of maintaining CHE and producing a GRB associated with a hypernova (\citealt{Nomoto05}) is not excluded.

Some other effects such as shorter lifetimes or wind anisotropies can
decrease the angular momentum loss and help maintain the angular momentum (\citealt{Meynet07}).
But even for a large helium star with sufficient angular momentum, there is
still an extra problem for producing a GRB. A relativistic jet generated in the stellar center cannot break out of the star if the
duration of central engine is shorter than the jet crossing time
inside the star (\citealt{Meszaros01}). Taking a radius of
$\sim10^{12}$ cm for a helium star with a mass $\sim100M_{\odot}$, the jet crossing
time inside the star is estimated as $t_{\rm cross}\gtrsim
100(r_{\rm H_{e}}/10^{12}$ cm) s (\citealt{Meszaros01}), which
requires the duration of GRB central engine to be longer than 100
seconds. Such durations are observed, but rare, compared to those of durations $<100$ s
(\citealt{Sakamoto11}).

\section{Fate of Dusty Clouds and Possible Dark Gamma-Ray Burst?}\label{Dark}

What is the fate of the dusty clouds near VFTS 682 after its death?
Will the clouds be destroyed by the explosion of VFTS 682? Because the
properties of dusty clouds around VFTS 682 are not at all unusual in
the 30 Dor region, we adopt the gas density of the 30
Dor core region $n\sim200$ cm$^{-3}$ (\citealt{Kawada11}) with a typical value
of $N_{\rm H}/A_{V}\sim0.7\times10^{22}$ cm$^{-2}$ in the LMC (\citealt{Schady07}). This implies that the foreground
extinction region has a size of $\sim50$ pc for $A_{V}\approx4.45$. This is consistent
with the 8.0 $\mu$m image in Figure \ref{fig1}, where the dusty region
around VFTS 682 has a projected size of $\sim30-40$ pc. In the ``walkaway'' scenario, VFTS 682 has both a tangential and RV velocity of
$\sim$30 km s$^{-1}$ away from R136. Is so, VFTS 682 will probably escape the dusty region in
its remaining $\sim2$ Myr lifetime. On the other hand, if VFTS
682 formed in situ, or has a much shorter lifetime, it will be still in the very dusty clouds at its death.

Since the luminosity of the afterglow of a GRB cannot
be well determined from just the progenitor mass and rotation, it is uncertain whether the potential GRB from VFTS 682 will be
dark, or the dusty clouds will be destroyed by the optical flashes produced by VFTS 682.
Typically we would assume a relativistic
jet breaks out of the star and emits an isotropic gamma-ray luminosity
$L_{\gamma}^{\rm iso}\sim10^{51}$ erg s$^{-1}$ (\citealt{Lee00}), followed by an early
optical afterglow with a luminosity $L_{\rm opt}^{\rm iso}\simeq0.1L_{\gamma}^{\rm iso}$.
In this case any dust region smaller than $R\sim30(L_{\rm opt}^{\rm iso}/10^{50}$ erg s$^{-1})^{1/2}$ pc cannot
survive because of dust sublimation by the optical flash and
fragmentation by the burst and afterglow (\citealt{WD00}; \citealt{Reichart02}).
In other words, the dusty region covering VFTS 682 and R136 would probably be
destroyed due to an early optical afterglow $L_{\rm opt}^{\rm iso}>10^{50}$ erg
s$^{-1}$. Otherwise, the less luminous early X-ray/optical afterglows will
heat and ionize the surrounding environment out to $\sim$ 100 pc, decreasing the dust column density
on a timescale of tens to hundreds of minutes in the observer's frame (\citealt{Perna98}).

However, it will be another story if VFTS 682 produces an intrinsically faint intrinsic faint GRB.
It is possible that the inefficient jet propagation inside the star only gives an underluminous
burst, as well as a dim optical afterglow ($L_{\rm opt}^{\rm iso}<10^{50}$ erg
s$^{-1}$) linked with the
weak burst. Then the dusty clouds will be heated and ionized,
but not be totally destroyed.

\section{Conclusions and Discussion}\label{Con}

We give a first order estimate whether VFTS 682 can be a LGRB progenitor in the sense of CHE, which is still a question mark in previous work. 
VFTS 682 will most likely lose more than 85\% of its current angular
momentum and cease its CHE evolution before it leaves MS with a remaining mass
$\sim50-70M_{\odot}$. In general VFTS
682 will fail to produce a GRB due to its strong mass-loss and
the resulting angular momentum loss, unless it has a currently extreme
rapid rotation ($v_{0}>700$ km s$^{-1}$) which helps it maintain CHE and
eventually produce a GRB associated with a hypernova. Wind anisotropies
and shorter lifetime could leave a more massive faster rotating star, but it is doubtful
whether a relativistic jet could travel through such a thick stellar
envelope and break out of the stellar surface, unless it is a rare long-lived ($\gtrsim100$ s) LGRBs.
Similarly, it is unlikely that VFTS 682 will be heavily obscured at death and produce a dark GRB. Its proper motion
will probably cause it out of the dusty region, and a GRB of an early optical afterglow
$L_{\rm opt}>10^{50}$ erg s$^{-1}$ would destroy the dust clouds within 30 pc,
otherwise the dusty clouds can be heated and ionized up to a region of $\sim$ 100 pc
by X-ray and optical radiation given by the death of VFTS 682.

CHE is believed to be the crucial process in the evolution path
towards LGRBs. However, the observed sample of WRs likely undergoing CHE is
quite small. The only observation of WR stars besides VFTS 682 which
might be undergoing are two WNh stars in the SMC (i.e.,
SMC-WR1 and WR2 in \citealt{Martins09}), which makes these three WR stars
extremely important to understanding the evolution of WR stars undergoing CHE
and the related problem of LGRB formation. The angular momentum losses
of the two WR stars in the SMC will be much less significant than VFTS
682 in the LMC. Taking the stellar parameters in
\cite{Martins09}, we estimate that the nuclear timescale of SMC-WR1 (WR2)
$\tau_{\rm nuc}\sim4$ Myr (5 Myr) is shorter than the wind timescale $\tau_{\rm wind}\sim10-20$
Myr, and the final stellar rotation velocity should be 90\% of the current
velocity. Therefore, the two WR stars in the SMC are more likely GRB progenitors in the scenario of CHE, although the metallicity
threshold is still an issue (\citealt{Martins09}). In any case, finding CHE WR stars
in the LMC and SMC promotes a future work on theory models.

VFTS 682 is a very interesting star. As mentioned in
Section 2, because there is no observation evidence to show that VFTS
682 is unusual compared to other stars in the 30 Dor region, the
interesting possibility of existence of others massive stars like VFTS
682 in the vicinity of R136 is not excluded. Note that R136 is
sufficient young and massive ($\leq5.5\times10^{4}M_{\odot}$) to
generate runaway stars beyond $150M_{\odot}$. Recently
\cite{Banerjee11} gives a theoretical model study on the
dynamical ejection of runaway massive stars from R136. We suggest that
there might be other massive stars that ``walked away'' from R136, but
are currently hidden behind even more dust than VFTS 682. Since mid-IR date cannot be used to flag such stars, spectroscopic observations, which are beyond the scope of this Letter, should be further investigated to show the possibility of their existence. 

\acknowledgments

We are grateful to C. S. Kochanek, T. A. Thompson and S. Horiuchi for useful comments and critical reading of the manuscript, and M. Pinsonneault, R. Khan, J. Beacom and J. Skowron for helpful discussions. This work is partially supported by the NSF grant AST-1108687.

\end{document}